\documentstyle[
aps, 
epsfig, 
tighten,
preprint,
]{revtex}
\newcommand{\beq}{\begin{equation}}
\newcommand{\eeq}{\end{equation}}
\newcommand{\beqa}{\begin{eqnarray}}
\newcommand{\eeqa}{\end{eqnarray}}
\newcommand{\Psib}{\overline{\Psi}}

\newcommand{\Lcal}{{\cal L}}
\newcommand{\p}{\partial}
\newcommand{\se}{\Sigma}
\newcommand{\bm}[1]{\mbox{\protect\boldmath $#1$}}

\begin{document}               
\draft


\title{Application of the density dependent hadron field theory to neutron star
matter}
\author{F. Hofmann, C. M. Keil, H. Lenske}
\address{Institut f\"ur Theoretische Physik, Universit\"at Gie\ss en,
         Heinrich-Buff-Ring 16, 35392 Gie\ss en, Germany}
\date{\today}

\maketitle

\begin{abstract}                

The density dependent hadron field (DDRH) theory, previously applied to isospin
nuclei and hypernuclei is used to describe $\beta$-stable matter and neutron
stars under consideration of the complete baryon octet. The meson-hyperon
vertices are derived from Dirac-Brueckner calculations of nuclear matter and
extended to hyperons. We examine properties of density dependent interactions
derived from the Bonn A and from the Groningen NN potential as well as
phenomenological interactions. The consistent treatment of the density
dependence introduces rearrangement terms in the expression for the baryon
chemical potential. This leads to a more complex condition for the 
$\beta$-equilibrium compared to standard relativistic mean field (RMF) approaches. 
We find a strong dependence of the equation of state and the particle 
distribution on the choice of the vertex density dependence. Results for neutron
star masses and radii are presented. We find a good agreement with other models 
for the maximum mass. Radii are smaller compared to RMF models and indicate a 
closer agreement with results of non-relativistic Brueckner calculations.

\end{abstract}

\pacs{PACS number(s): 26.60.+c, 21.65.+f, 21.30.Fe, 97.60.Jd}


\section{Introduction}
\label{sec:Intro}

The nuclear equation of state (EoS) is the fundamental input for the calculation
of neutron star properties. The internal structure of a neutron star ranges from
sub-nuclear densities at the surface to a few times the normal nuclear matter
density $\rho_0$ in its core. Therefore, a detailed knowlegde of the EoS over a
wide range of densities is required. Of particular importance is the behavior of
the EoS for densities $\rho \gg \rho_0$ since it primarily determines the
maximum mass of the star. Comparision with experimentally observed data for
neutron star masses and radii might allow us to examine the reliability of
different models for the EoS at high densities.

The theoretical determination of such an EoS is a very hard task since there is
only accurate knowledge around $\rho=\rho_0$ where matter in the weak
equilibrium consists mainly of nucleons and leptons. At higher densities of $2-
3\,\rho_0$ one expects strange baryons to appear as new hadronic degrees of
freedom. The extrapolation of the EoS to high densities and its implication on
neutron star properties has been examined in a variety of models, most notably
in the framework of phenomenological non-relativistic potential models
\cite{Balberg:97}, relativistic mean-field approaches (RMF) 
\cite{Glendenning:82,Weber:89a,Knorren:95,Schaffner:96} 
and non-relativistic Brueckner-Hartree-Fock (BHF) 
calculations \cite{Baldo:00,Vidana:00}. In addition there have been
calculations employing chiral effective Lagrangians \cite{Hanauske:00} or the
quark meson coupling model \cite{Pal:99}. From a microscopic point of view
Brueckner calculations using realistic nucleon-nucleon (NN) and hyperon-nucleon
(YN) potentials as input have to be preferred. Since the density inside neutrons
stars is extremely high - the Fermi momenta and the baryon effective mass are in
the order of $500$ MeV - the use of non-relativistic models might be problematic
and one should prefer a relativistic description. Due to technical difficulties,
relativistic (Dirac) BHF calculations are restricted to asymmetric nuclear
matter and the inclusion of hyperons at higher densities is at present not
feasible. Therefore, only DB calculations for neutron star matter with nucleons
have been performed \cite{Muether:87,Engvik:94}.

RMF theory \cite{Walecka:74,Serot:86} allows to incorporate easily and
consistently an enlarged set of hadronic degrees of freedom and has been
succesfully applied to hypernuclei 
\cite{Rufa:87,Mares:89,Rufa:90,Glendenning:93} 
and neutron stars \cite{Glendenning:82,Knorren:95,Schaffner:96}. Being a
phenomenological model with its parameters usually adjusted to the properties of
finite nuclei and their limited density range around $\rho_0$, an extrapolation
to higher densities has some uncertainties. Parameterizations with an excellent
describtion of finite nuclei turn out to be unstable at higher densities being
mainly caused by divergent scalar self-interaction terms \cite{Reinhard:88}.
Nucleons aquire negative effective masses and the equation of state turns out to
be much stiffer compared to Dirac-Brueckner calculations \cite{Schaffner:96}.
One solution is the introduction of a quartic vector self-interaction term
\cite{Bodmer:91,Sugahara:94}. Also, calculations in relativistic Hartree-Fock
approximation with coupling constants fitted to the EoS of DB calculations were
succesfully applied to neutron star matter \cite{Weber:89,Huber:98}.

In this paper we employ results from DB calculations by parameterizing the DB
self-energies in terms of density dependent coupling functionals and apply them
in Hartree approximation to strange and neutron star matter. This approach, the
density dependent hadron field (DDRH) field theory, has been introduced in
\cite{Brockmann:92,Lenske:95} and applied to stable 
\cite{Fuchs:95,Ineichen:95,Cescato:98} 
and exotic nuclei \cite{Hofmann:00a}. Taking into account
information from realistic NN potentials and a much wider density range ($0.5-
3\,\rho_0$) we expect our extrapolation to high densities to be more stable than
purely phenomenological RMF models.

An important point is the treatment of the hyperons. In the DDRH model they can
be treated in almost the same manner as in standard RMF theory, allowing an easy
extension of the model to the SU(3)$_f$ octet baryons as was shown in
\cite{Keil:00}. Since we assume in general density dependent hyperon-meson
vertices we should in principle derive their parameterization from DB
calculations that include hyperons. Since such calculations are not available we
extrapolate the density dependence and the strength of the vertices from DB
calculations of nuclear matter and from hypernuclear data. As discussed in
\cite{Keil:00} this provides at least qualitative information about DB results
of strange matter. This approach is discussed in Sec.~\ref{sec:DensDepHypMat}
where also a short review of the theoretical model is given. 
In Sec.~\ref{sec:BetaMat} 
we apply the derived density dependent hyperon-nucleon interaction to
$N \Lambda$ matter and to neutron star matter in $\beta$-equilibrium. We discuss
the influence of the density dependence on the chemical potential and present
results for the equation of state for various density dependent interactions.
Neutron star matter compositions for different models are presented. In
Sec.~\ref{sec:NeuStars} the influence of the calculated EoS on mass-radius
relations of neutron stars is investigated and compared with previous results of
other groups. The paper closes in Sec.~\ref{sec:Summary} with a summary and
conclusions.

\section{Density dependent hadron field theory with hyperons}
\label{sec:DensDepHypMat}
\subsection{The Model Lagrangian}
\label{ssec:Lagr}

The model Lagrangian closely follows in structure the one of relativistic mean-
field (RMF) theory \cite{Walecka:74,Serot:86}. The important difference in the
density dependent relativistic hadron field theory (DDRH) corresponds to
replacing the constant meson-baryon vertices of the RMF model by functionals
$\hat \Gamma(\hat \rho)$ of Lorenz-scalar bilinear forms $\hat \rho(\Psib_F,
\Psi_F)$ of the baryon field operators. This step is necessary to retain
thermodynamical consistency and energy-momentum conservation for this extended
model and has been discussed in detail in \cite{Lenske:95,Fuchs:95}. In the
meson-exchange particle sector we have also included the scalar isovector meson
$\delta$ in contrast to previous formulations since it is expected to be
important in very asymmetric systems.

In this work the extension to hypernuclei as introduced in \cite{Keil:00} is
used. The $1/2^+$ baryon octet is taken into account including the $S=-1\,
(\Lambda, \Sigma)$ and $S=-2\,(\Xi)$ hyperons. Besides the standard set of non-
strange mesons we also include the hidden-strangeness meson fields $\sigma_s$
(scalar, $m_{\sigma_s}=975$ MeV) and $\phi$ (vector, $m_{\phi}=1020$ MeV)
\cite{Schaffner:94}.

Defining the flavour spinor $\Psi_F$
\beq
\Psi_F = \left( \Psi_N, \Psi_\Lambda, \Psi_\Sigma, \Psi_\Xi \right)^{\text{T}}
\label{eq:FlSp}
\eeq
which is composed of the isospin multiplets
\beqa 
\Psi_N = \left( \begin{array}{c} \psi_p \\ \psi_n \end{array} \right) &,&\quad
\Psi_\Lambda = \psi_\Lambda, \nonumber \\ 
\Psi_\Sigma = \left( \begin{array}{c} \psi_{\Sigma^+} \\
\psi_{\Sigma^0} \\ \psi_{\Sigma^-} \end{array} \right) &,&\quad
\Psi_\Xi = \left( \begin{array}{c} \psi_{\Xi^0} \\  \psi_{\Xi^-} \end{array}
\right),
\eeqa
the Lagrangian is written as
\beqa
\Lcal &=& \Lcal_{B} + \Lcal_{M} + \Lcal_{int} \nonumber \\
\Lcal_{B} &=& \Psib_F \left[ i\gamma_\mu\p^\mu - \hat M_F \right] \Psi_F \\
\Lcal_{M} &=&\frac{1}{2} \sum_{i=\sigma,\delta,\sigma_s}
    \left(\p_\mu\Phi_i\p^\mu\Phi_i - m_i^2\Phi_i^2\right) - \nonumber \\
    && \frac{1}{2} \sum_{\kappa=\omega,\rho,\phi}
    \left( \frac{1}{2} F^{(\kappa)}_{\mu\nu} F^{(\kappa)\mu\nu} - m_\kappa^2
    A^{(\kappa)}_\mu A^{(\kappa)\mu}
    \right) \label{eq:Lagrangian} \\
\Lcal_{int} &=&
     \Psib_F\tilde{\Gamma}_{\sigma}  \Psi_F\Phi_{\sigma}
 -   \Psib_F\tilde{\Gamma}_{\omega}  \gamma_{\mu}\Psi_F A^{(\omega)\mu} +
 \nonumber \\
&&   \Psib_F\tilde{\Gamma}_{\delta}  \bm{\tau}\Psi_F\bm{\Phi}_{\delta}
 -   \Psib_F\tilde{\Gamma}_{\rho}    \gamma_{\mu}\bm{\tau}\Psi_F\bm{A}^{(\rho)
 \mu} + \nonumber \\
&&   \Psib_F\tilde{\Gamma}_{\sigma_s}\Psi_F\Phi_{\sigma_s}
 -   \Psib_F\tilde{\Gamma}_{\phi}    \gamma_{\mu}\Psi_F A^{(\phi)\mu}.
\eeqa

$\Lcal_B$ and $\Lcal_M$ are the free baryonic and mesonic Lagrangians,
respectively. Baryon-meson  interactions are described by $\Lcal_{int}$ that
includes the vertex functionals $\tilde{\Gamma}(\Psib_F,\Psi_F)$. The diagonal
matrix $\tilde M$ contains the free-space baryon masses. The Lagrangian and the
interaction have to be symmetric in flavor space, since SU(3)$_f$-flavor
exchanging mesons are not considered. This is obtained by defining the vertices
as
\beq
\left( \tilde{\Gamma}_\alpha \right)_{BB'} = \hat\Gamma_{\alpha B}\delta_{BB'}.
\eeq
The indices $\alpha = \sigma,\omega,\delta,\rho,\sigma_s,\phi$ and $B = N,
\Lambda,\Sigma,\Xi$ denote all mesons and baryon multiplets. The most general
\emph{ansatz} for the DDRH vertices, allowing to treat the density dependence of
each
vertex independently, is
\beqa
\hat\Gamma_{\alpha B}\left(\hat{\rho}_{\alpha B}(\Psib_F, \Psi_F)\right)
\eeqa
where $\hat{\rho}_{\alpha B}$ is a Lorentz-scalar combination of the baryon
field operators.
As discussed in the introduction, the strength and the intrinsic density
dependence of the vertices have to be deduced from microscopic calculations. The
mapping of the DBHF self-energies to infinite nuclear matter DDRH vertices is
done in the local density approximation (LDA), 
e.g.~\cite{Brockmann:92,Haddad:93,deJong:98a}, 
and has been thoroughly discussed in \cite{Hofmann:00a,Keil:00}.
In \cite{Hofmann:00a} a momentum correction of the self-energies was introduced
to improve the applicability of DB calculations to the relativistic Hartree
approximation. DB self-energies derived in asymmetric nuclear matter
\cite{deJong:98a} were parameterized and special care was taken to reproduce the
DB binding energy over the complete asymmetry range from isospin symmetric
nuclear matter to pure neutron matter.

The standard choice is to let the density operator $\hat \rho$ depend on the
baryon vector current $\hat j_\mu = \Psib \gamma_\mu \Psi$. This so called
vector density dependence (VDD) leads to quite satisfactory results for finite
nuclei \cite{Fuchs:95} and hypernuclei \cite{Keil:00} and is a natural choice
for the parameterization of the DB self-energies as was discussed in
\cite{Hofmann:00a}. A straightforward extension to meson-hyperon vertices is the
\emph{ansatz} \cite{Keil:00}
\beq
\hat{\rho}_{\alpha B}[\Psib_F, \Psi_F] = \Psib_F
\tilde{B}^\mu_{\alpha B}\gamma_\mu \Psi_F.
\label{eq:RhoNod}
\eeq
Different choices of the matrix $\tilde{B}^\mu_{\alpha B}$ and their physical
significance will be discussed in the next section. Taking the variational
derivative of the Lagrangian
\beq \label{eq:varL}
 \frac{\delta\Lcal_{int}}{\delta\Psib_F} = \frac{\p\Lcal_{int}}{\p\Psib_F} +
\sum_{\alpha,B}
\frac{\p\Lcal_{int}}{\p\hat{\rho}_{\alpha B}} \frac{\delta\hat{\rho}_{\alpha
B}}{\delta\Psib_F}.
\eeq
leads to the usual meson field equations known from RMF theory but replacing the
constant couplings $g_{\alpha B}$ by density dependent vertices
$\hat\Gamma_{\alpha B}(\hat{\rho})$. In the baryon field equations additional
rearrangement contributions appear that have their origin in the second term of
Eq.~(\ref{eq:varL}),
\beq
\left[\gamma_\mu \left( i\p^\mu - \tilde\se^{\mu(0)} - \tilde\se^{\mu(r)}
\right) - \left(
\tilde M - \tilde\se^{s(0)} \right) \right] \Psi_F = 0.
\eeq

The scalar self-energy $\tilde \se^{s(0)}$ has the standard form, while the
vector self-energy $\tilde \se^{\mu} = \tilde\se^{\mu(0)} + \tilde\se^{\mu(r)}$
includes rearrangement contributions $\tilde \se^{\mu(r)}$ introduced by the
medium-dependence of the vertices.
\beqa \label{eq:self_s}
\tilde\se^{s(0)}   & = & \tilde\Gamma_{\sigma}\Phi_{\sigma} +
                         \tilde\Gamma_{\delta}\bm{\tau}\bm{\Phi}_{\delta} +
                         \tilde\Gamma_{\sigma_s}\Phi_{\sigma_s} \\ \label{eq:
                         self_v}
\tilde\se^{\mu(0)} & = & \tilde\Gamma_{\omega}A^{(\omega)\mu} +
                         \tilde\Gamma_{\rho}  \bm{\tau}\bm{A}^{(\rho)\mu} +
                         \tilde\Gamma_{\phi}  A^{(\phi)\mu} \\ \label{eq:self_r}
\tilde\se^{\mu(r)} & = & \sum_{\alpha,B} \frac{\p\Lcal_{int}}{\p\hat{\rho}
_{\alpha B}}
                         \tilde B_{\alpha B}^{\mu}
\eeqa

The importance of the rearrangement energies $\tilde\se^{\mu(r)}$ and their
physical origin has been discussed in detail in \cite{Fuchs:95,Negele:82}.
Their explicit form and their impact on neutron star matter will be discussed in
the next sections.

\subsection{Choice of the density dependence}
\label{ssec:DensDep}

The density dependence of the hyperon-meson vertex functionals should in
principle be derived directly from DB self-energies of strange matter performed
with the complete baryon octet. However, such a full scale calculation is not
available and hardly feasible under present conditions. Non-relativistic
Brueckner calculations with microscopic interactions have been performed for
strange matter by several groups \cite{Schulze:98,Stoks:99,Vidana:00b} but,
obviously, their results can not be used in relativistic calculations because of
the different structure of the non-relativistic single particle potentials.
Therefore, we choose a semiempirical approach by relating the microscopic
density dependence of the meson-nucleon vertices derived from DB calculations of
pure isospin matter (only $p,n$) to the density dependence of the hyperon-
nucleon and hyperon- hyperon interaction.

In \cite{Keil:00} we have shown by inspecting the properties of the Dirac-
Brueckner interaction in strange matter that density dependent nucleon and
hyperon dynamics can be related to each other to a good approximation by scaling
laws. The main outcome is that the hyperon and nucleon self-energies and
vertices are related to each other by the ratio of the free space coupling
constants $\bar g_{\alpha B}$. In leading order Hartree approximation one finds
the relation
\beq
\label{eq:scaleGamma}
R_{\alpha Y} = \frac{\Gamma_{\alpha Y}}{\Gamma_{\alpha N}}
\simeq \frac{\se_{\alpha Y}}{\se_{\alpha N}}
\simeq \frac{\bar g_{\alpha Y}}{\bar g_{\alpha N}}.
\eeq

Since this relation is strictly valid only for symmetric hypermatter with the
same content of nucleons and hyperons \cite{Keil:00} it is not obvious how the
density dependence of the hyperon vertices has to be related to the full
baryonic density for other matter compositions. An obvious and rather simple
approach is to let all vertices depend on the total baryon density, denoted by
$\hat \rho_T$. From now on we will denote this choice as model 1. For the
hyperons this corresponds to defining the coupling functionals as
\beqa
\hat\Gamma_{\alpha\Lambda} & = & R_{\alpha\Lambda} \hat\Gamma_{\alpha N}(\hat
\rho_T), \quad \nonumber \\
\hat\Gamma_{\alpha\Sigma}  & = & R_{\alpha\Sigma}  \hat\Gamma_{\alpha N}(\hat
\rho_T),  \quad \nonumber\\
\hat\Gamma_{\alpha\Xi}     & = & R_{\alpha\Xi}     \hat\Gamma_{\alpha N}(\hat
\rho_T).
\eeqa
The parameterization of the density dependence is taken from the nucleon-
vertices
$\hat\Gamma_{\alpha N}(\hat \rho_{N})$ derived from DB calculations of nuclear
matter where
$\hat \rho_N \equiv \hat \rho_T$. By relating the density $\hat \rho_T$ to the
total baryon vector current
\beq \hat \rho_T =
\sqrt{\hat j_{\mu}^F \hat j^{\mu F}}.
\eeq
a Lorentz-invariant expression is obtained which is equivalent to choosing the
matrix $\tilde B_\mu^{\alpha B}$ in Eq.~(\ref{eq:RhoNod}) as
\beq
\tilde B_\mu^{\alpha B} \equiv \tilde B_{\mu} = \hat u_{\mu} \text{diag} \left(
1,1,1,1 \right),
\eeq
where $\hat u^{\mu}$ is a four velocity with $\hat u^{\mu}\hat u_{\mu}=1$. 
This choice leads to rearrangement self-energies of the form
\beqa
\hat\se^{\mu(r)} & = & \sum_{B'} \left(
   \frac{\p\hat\Gamma_{\omega B'}}  {\p\hat\rho_{B'}}A^{(\omega)}_{\nu}\hat
   j^{\nu}_{B'}
 - \frac{\p\hat\Gamma_{\sigma B'}}  {\p\hat\rho_{B'}}\Phi_{\sigma}\hat
 \rho^s_{B'}
 \right. \nonumber \\ & & \left.
 + \frac{\p\hat\Gamma_{\rho B'}}    {\p\hat\rho_{B'}}\bm{A}^{(\rho)}_{\nu}
 \Psib_{B'} \gamma^\nu \bm{\tau}\Psi_{B'}
  - \frac{\p\hat\Gamma_{\delta B'}}  {\p\hat\rho_{B'}}\bm{\Phi}_{\delta}
 \Psib_{B'} \bm{\tau}\Psi_{B'}
 \right. \nonumber \\ & & \left.
 + \frac{\p\hat\Gamma_{\phi B'}}    {\p\hat\rho_{B'}}A^{(\phi)}_{\nu}\hat
 j^{\nu}_{B'}
 - \frac{\p\hat\Gamma_{\sigma_s B'}}{\p\hat\rho_{B'}}\Phi_{\sigma_s}\hat
 \rho^s_{B'}
   \right) \hat u^{\mu}
\label{eq:rearr1}
\eeqa
where the scalar densities are defined as $\hat \rho^s_{B} = \Psib_B \Psi_B$.
Since the density dependence of all vertices was chosen to depend on the same
argument $\hat \rho_T$ the summation has to be performed over all $B'$ leading
to identical rearrangement self-energies for all baryon multiplets. Also, in a
system consisting of different fractions of hyperons and nucleons, the scaling
relation for the vertices is fulfilled for all densities and independent of the
strangeness fraction.

In Ref.~\cite{Keil:00} $\Lambda$-hypernuclei were considered and the medium
modification of the vertices was chosen to depend only on the density of the
surrounding baryons of same particle type, e.g. $\hat\Gamma_{\alpha \Lambda}
(\hat \rho_{\Lambda})$. The parameterization of the density dependence was taken
from the nucleon-vertices $\hat\Gamma_{\alpha N}(\hat \rho_{N})$ derived from
DB calculations of nuclear matter. We will denote this choice as model 2.
Extending this model to the complete baryon octet, we let the vertices only
depend on the density of baryons of the same multiplet
\beqa
\hat\Gamma_{\alpha\Lambda} & = & R_{\alpha\Lambda} \hat\Gamma_{\alpha N}(\hat
\rho_{\Lambda}), \quad \nonumber \\
\hat\Gamma_{\alpha\Sigma}  & = & R_{\alpha\Sigma}  \hat\Gamma_{\alpha N}(\hat
\rho_{\Sigma}),   \quad \nonumber \\
\hat\Gamma_{\alpha\Xi}     & = & R_{\alpha\Xi}     \hat\Gamma_{\alpha N}(\hat
\rho_{\Xi}),
\eeqa
where the densities are the baryon vector currents of the corresponding baryon
species
\beq
\hat \rho_{B} = \sqrt{\hat j_{\mu}^B \hat j^{\mu B}}.
\eeq
In this model the matrix $\tilde B_\mu^{\alpha B}$ in Eq.~(\ref{eq:RhoNod}) is
defined as
\beq
\tilde B_\mu^{\alpha B} \equiv \tilde B_{\mu}^{B} = \hat u_{\mu} \text{diag}
\left(
\delta^{NB},\delta^{\Lambda B},\delta^{\Sigma B},\delta^{\Xi B} \right)
\eeq
and the rearrangement self-energies now differ between the baryon multiplets
\beqa
\hat\se^{\mu(r)}_B & = &  \left(
   \frac{\p\hat\Gamma_{\omega B}}  {\p\hat\rho_B}A^{(\omega)}_{\nu}\hat j^{\nu}
   _B
 - \frac{\p\hat\Gamma_{\sigma B}}  {\p\hat\rho_B}\Phi_{\sigma}\hat \rho_B^s
\right. \nonumber \\ & & \left.
 + \frac{\p\hat\Gamma_{\rho B}}    {\p\hat\rho_B}\bm{A}^{(\rho)}_{\nu} 
 \Psib_B \gamma^\nu \bm{\tau}\Psi_B
 - \frac{\p\hat\Gamma_{\delta B}}  {\p\hat\rho_B}\bm{\Phi}_{\delta}
 \Psib_B \bm{\tau}\Psi_B
\right. \nonumber \\ & & \left.
 + \frac{\p\hat\Gamma_{\phi B}}    {\p\hat\rho_B}A^{(\phi)}_{\nu}\hat j^{\nu}_B
 - \frac{\p\hat\Gamma_{\sigma_s B}}{\p\hat\rho_B}\Phi_{\sigma_s}\hat \rho_B^s
   \right) \hat u^{\mu}.
\label{eq:rearr2}
\eeqa
This separation of the density dependence also corresponds in first order to the
outcome of DB
considerations and leads to satisfying results for $\Lambda$-hypernuclei. It
takes into account that in leading order the medium dependence of the vertices
is only caused by Pauli-blocked intermediate states of baryons of the same
multiplet. On the other hand, this separation leads to a variation of the
relative strength of the vertices in strongly asymmetric systems, e.g. neutron
stars, since the vertices depend on different arguments $\hat\rho_B$. This
behavior is different compared to standard RMF calculations where the ratio of
the vertices stays constant over the whole density range and is independent of
the strangeness content.

The two vertex models constitute limiting cases where model 1 (dependence on
$\rho_T$) is probably most realistic for systems at high total baryon densities.
Modell 2 (dependence on $\rho_B$), on the other hand, can be expected to be most
realistic for low hyperon densities as found in single $\Lambda$-hypernuclei. We
assume that a realistic density dependence should be composed as a mixture of
the total density and the baryon multiplet density
\beq
\hat\Gamma_{\alpha B} \equiv \hat\Gamma_{\alpha B}(\hat\rho_B,\hat\rho_T)
\eeq

Unfortunately, DB calculations from which such a functional dependence could be
extracted are not yet available. Our choice, taking into account both extremes,
allows us to examine thoroughly the properties of such a parameterization.
Furthermore, for pure systems, consisting only of e.g. $\Lambda$, both models
are identical.

\subsection{The vertex scaling factors}
\label{ssec:VertexScale}

In RMF theory the phenomenological hyperon and nucleon vertices can be related
to each other by simple scaling factors $R_{\alpha Y}$, e.g. $g_{\alpha \Lambda}
= R_{\alpha \Lambda}g_{\alpha N}$. As shown in Section \ref{ssec:DensDep} this
relation also holds in the DDRH model. A widely used approach is to determine
these vertex scaling factors from SU(6) symmetry relations of the quark model
\cite{Dover:85}. For the vector mesons ideal mixing is assumed and one finds
\cite{Schaffner:94}
\beqa
\Gamma_{\omega\Lambda} = \Gamma_{\omega\Sigma} &=& 2\Gamma_{\omega\Xi} =
\frac{2}{3}\Gamma_{\omega N}, \nonumber \\
\Gamma_{\rho\Sigma} = 2\Gamma_{\rho\Xi} &=& 2\Gamma_{\rho N}, \quad
\Gamma_{\rho\Lambda} = 0 \\
2\Gamma_{\phi\Lambda} = 2\Gamma_{\phi\Sigma} &=& 2\Gamma_{\phi\Xi}
= - \frac{2\sqrt{2}}{3}\Gamma_{\omega N} \nonumber \\ 
\Gamma_{\phi N} &=& 0.
\eeqa
The scalar coupling constants are fitted to hypernuclear properties 
\cite{Rufa:87,Mares:89,Rufa:90} 
and chosen to give hyperon potentials in saturated nuclear
matter that are compatible with experimental results for the single particle
spectra of hypernuclei.
\beq
U_{\Lambda}^{(N)} = U_{\Sigma}^{(N)} = -30 \text{ MeV}, \quad U_{\Xi}^{(N)} = -
28 \text{MeV}.
\eeq
The values of the $\Lambda$, $\Sigma$ and $\Xi$ potentials were chosen in
accordance with Ref. \cite{Schaffner:96,Schaffner:94}. Based on the analysis of
$\Sigma^-$ atomic data the real part of the optical potential was found to be
negative \cite{Dover:89}. A recent analysis, however, indicates that the
isoscalar potential changes sign in the nuclear interior and becomes repulsive
\cite{Mares:95}. Therefore, we also performed calculations with a potential
depth of $U_{\Sigma}^{(N)} = +30$ MeV and found that the $\Sigma$ does not
appear in the composition of neutron star matter. For this reason we restrict
our discussion to the negative potential value. The $\Xi$ nuclear interaction
also exhibits large uncertainties with potential depths ranging from $-14$ MeV
to $-28$ MeV in some investigations \cite{Schaffner:00}.

In DDRH theory the non-relativistic single particle potentials of hyperons in
nuclear matter are obtained as
\beq
U_{Y}^{(N)} = \se_{Y}^{0(0)} + \se_{Y}^{0(r)} - \se_{Y}^{s(0)}.
\eeq
Choosing $U_{Y}^{(N)}$ as fixed by phenomenology this relation introduces a
constraint on the couplings $\Gamma_{\sigma Y}$ and $\Gamma_{\omega Y}$.
Due to the different density dependence of model 1 and 2 one finds different
values for the scaling factors $R_{\sigma Y}$ even though they were adjusted to
the same potential depth.
In model 1 the hyperon vertices have to be evaluated at saturation density
$\rho_T=\rho_0$ even though the density of the hyperons is $\rho_Y\ll\rho_T$. In
addition, the rearrangement contributions of the nucleons add to the hyperon
potential. In contrast, in model 2 the vertices  have to be calculated at
vanishing hyperon density and rearrangement does not contribute to the hyperon
potential.

We examined the behavior of three different density dependent interactions. The
Bonn A parameter set \cite{Haddad:93} parameterizes the density dependence of DB
Brueckner calculations in symmetric nuclear matter with the Bonn A NN potential
\cite{Brockmann:90,Machleidt:89}. It only includes density dependent vertices
for the $\sigma$ and the $\omega$ meson while the $\rho$ meson coupling strength
was chosen as a constant. Calculations for single $\Lambda$-hypernuclei have
been performed successfully with this parameterization \cite{Keil:00}. The
Groningen parameter set \cite{Hofmann:00a} was fitted to DB Brueckner
calculations in asymmetric nuclear matter with the Groningen NN potential
\cite{deJong:98a,deJong:98b,Malfliet:88}. In addition to the isoscalar channel
it includes a density dependence in the isovector part of the interaction,
parameterized by the $\rho$ and the $\delta$ nucleon-meson vertices. The third
parameter set is a phenomenological interaction whose density dependence
($\sigma$, $\omega$ and $\rho$ meson) has been adjusted by a fit to nuclear
matter properties and finite nuclei \cite{Typel:99}. It will be denoted from now
on as DD. The quality of its description of finite nuclei is comparable to RMF
parameter sets including nonlinear meson interaction terms and its nuclear
matter properties at higher densities are in accordance with DB calculations.
Results for the sccaling factors are presented in the right column of
Tab.~\ref{tab:RsRw}. From now on, the set of scaling factors $R_{\alpha Y}$ that
was determined from the SU(6) symmetry relations of the quark model will be
denoted as $R_q$.

It is our aim to start from DB theory and use microscopic interactions as input
for our calculations. Since these results include highly nonlinear and
nonperturbative correlation effects, the strict use of the quark model reduction
of $R_{\omega \Lambda}=2/3$ might be questionable. Therefore, we examine in
addition the properties of microscopically derived scaling factors. For the Bonn
A potential an extension to the free $N\Lambda$ systems exist 
\cite{Reuber:96,Haidenbauer:98}, 
but DB calculations are pending. A free space scalar vertex
scaling factor $R_{\sigma\Lambda} = 0.49$ was extracted from $N\Lambda$ T-matrix
results for a sharp $\sigma$ meson mass $m_\sigma = 550$ MeV 
\cite{Haidenbauer:98}. 
Following Eq.~(\ref{eq:scaleGamma}) we can apply this value also to the in-
medium vertices.

Since microscopic values for $R_{\omega\Lambda}$ are not available from
Ref.~\cite{Haidenbauer:98} we use it as a phenomenological parameter and adjust
it to the potential depth $U_{\Lambda}^{(N)}$. In this analysis we are
restricted to the $\Lambda$ because calculations for  $N\Sigma$ and $N\Xi$
systems were not yet performed. For the Bonn A potential we find a value of
$R_{\omega\Lambda} = 0.569$ in model 2 in close agreement with \cite{Keil:00}
where a value of $R_{\omega\Lambda} = 0.553$ was found by a fit a $\Lambda$
single-particle energies. In a constant coupling RMF model \cite{Ma:96} a
relative $\omega$ coupling of $R_{\omega\Lambda} = 0.512$ was found for the same
value of $R_{\sigma\Lambda}$. This is again in close agreement with our
corresponding values of model 1, e.g. $R_{\omega\Lambda} = 0.510$ for the
phenomenological density dependence. Results for the vector meson scaling
factors $R_{\omega\Lambda}$ for model 1 and 2 for the different interactions are
shown in the left column of Tab.~\ref{tab:RsRw}. We will denote this set of
semi-microscopic scaling factors by $R_m$.

In \cite{Reuber:96,Haidenbauer:98} the scalar meson channels were described by
the correlated exchange of pion and kaon pairs. Therefore,the scalar coupling
also includes a relevant admixtures of the $\sigma_s$ field in the $\Lambda$
coupling. Adjusting the vector coupling we also implicitly
include contributions from the $\phi$ field. Naive quark counting suggests that
even for a pure $\Lambda$-$\Lambda$ interaction the strange mesons contribute
only about 10\% of the interaction strength. For this reason we neglect the
explicit contributions from the $\sigma_s$ and $\phi$ meson in our calculations.
In addition, it should be noted that the contribution from these hidden-
strangeness mesons can not be fixed by experimental data, introducing additional
ambiguities in the model as seen in \cite{Schaffner:00}.

In the next section we will show that there are already notable differences
between the different models without these additional fields. The notable
deviations of $R_{\sigma\Lambda}$ and $R_{\omega\Lambda}$ to the quark model
value of 2/3 originate in higher order nonlinear contributions from the
dynamically generated $\sigma$ and $\sigma_s$ exchange channels 
\cite{Reuber:96,Haidenbauer:98}, 
the explicit SU(3)$_f$ symmetry breaking and the $\omega$-
$\phi$ octet-singlet mixing on the fundamental strong interaction level.

\section{Mean-field description of strange and $\beta$-stable matter}
\label{sec:BetaMat}

\subsection{Properties of strange matter}
\label{ssec:StrangeMat}

We now apply the extended DDRH model of Sec.~\ref{sec:DensDepHypMat} to
$\Lambda$ matter and $\beta$-stable matter. A solvable scheme is obtained in
mean-field theory which amounts to taking the expectation values with respect to
the Hartree ground state of the baryon-meson vertices and the meson fields. The
vertices then reduce to density dependent functions $\Gamma_{\alpha B}(\rho)$ of
the baryon density $\rho$. In static infinite matter all derivative terms vanish
and the solution of the fields can be expressed analytically \cite{Serot:86}.

In weak $\beta$-equilibrium electrons $e^-$ and myons $\mu^-$ have to be
included in the
Lagrangian due to the weak decay channel,
\beq
\Psi_L = \left( \psi_e, \psi_{\mu} \right)^{\text{T}} \eeq \beq \Lcal_{L}=
\Psib_L
\left[ i\gamma_\mu\p^\mu - \tilde M_L \right] \Psi_L.
\eeq
In mean-field theory the energy density $\epsilon$ and the pressure $P$ are
given by the ground state expectation values of the energy-momentum tensor
$T^{\mu\nu}$. One finds
\beqa
\epsilon = \langle T^{00} \rangle & = &
    \sum_{i=b,l} \frac{1}{4} \left[ 3E_{F_i}\rho_i + m_i^*\rho_i^s \right] \\
    \nonumber & + &
    \sum_{b} \frac{1}{2} \left[ \rho_b\se_b^{0(0)} + \rho_b^s\se_b^{s(0)}
    \right]
\label{eq:eNucMat} \\
P = \frac{1}{3}\sum_{i=1}^3\langle T^{ii} \rangle & = &
    \sum_{i=b,l} \frac{1}{4} \left[ E_{F_i}\rho_i - m_i^*\rho_i^s \right] \\
    \nonumber & + &
    \sum_{b} \frac{1}{2} \left[\rho_b\se_b^{0(0)} - \rho_b^s\se_b^{s(0)} +
    2\rho_b\se_b^{0(r)} \right]
\label{eq:pNucMat}
\eeqa
where the indices $b=p,n,\Lambda,\Sigma^-,\Sigma^0,\Sigma^+,\Xi^-,\Xi^0$ and $l=
e^-,\mu^-$ run over all baryons and leptons, respectively. The self-energies are
the mean-field expectation values of the quantities defined in 
Eqs.~(\ref{eq:self_s})-(\ref{eq:self_r}), 
$E_{F_b}= \sqrt{k_{F_b}^2+{m_b^*}^2}$ is the energy
of the particle, $k_F$ is the Fermi momentum and $m^*_b = M_b - \se_b^s$ the
effective mass.

The chemical potential in a system with fixed baryon number is defined as
\beq
\label{eq:chem1}
\mu_b = \frac{\p\epsilon}{\p\rho_b} = \sqrt{k^2_{F_b} + {m_b^*}^2} +
\se_b^{0(0)} + \se_b^{0(r)}
\eeq
In contrast to constant coupling RMF models the rearrangement energy appears in
the above relation which is mandatory for thermodynamical consistency. This can
be easily verified by applying the Hugenholtz-van Hove theorem 
\cite{Hugenholtz:58} 
that relates energy and pressure of a particle to its chemical potential and
noting that the rearrangement contribution also appears in the expression for
the pressure.

In weak $\beta$-equilibrium the chemical potentials of all particles are related
to each other by
\beq
\label{eq:chem2}
\mu_i = b_i \mu_{n} - q_i \mu_{e}
\eeq
which imposes charge and baryon number conservation. Here, the index $i$ denotes
all particle species (baryons and leptons) and $b_i$ and $q_i$ are the
corresponding baryon number and electrical charge. The equilibrium problem is
solved by determining the meson fields from the field equations of the form
\beq
m_{\alpha}^2\Phi_{\alpha} = \sum_b \Gamma_{\alpha b}(\tilde \rho)\tilde \rho_b
\eeq
where $\tilde \rho$ depends on the choice of model 1 or 2. 
The densities $\tilde \rho_b$ are the (iso)scalar or (iso)vector densities and
are determined by the Lorentz and isospin structure of the corresponding meson
vertex. As an additional constraint the total baryon density $\rho$ is fixed and
charge neutrality is imposed for neutron star matter
\beq
\label{eq:constraint}
\rho = \sum_i b_i \rho_i, \quad \rho_c = \sum_i q_i \rho_i = 0.
\eeq
This allows the determination of the electron $\mu_{e}$ and baryon $\mu_{n}$
chemical potentials in $\beta$-equilibrium. In contrast to standard RMF
calculations one can not simply extract the Fermi momentum $k_{F_i}$ and thus
the density $\rho_i = k_{F_i}^3/3\pi^2$ of each particle for a fixed chemical
potential from Eqs.~(\ref{eq:chem1}) and (\ref{eq:chem2}). Since the self-
energies depend by virtue of the vertices on the densities of the baryons a
coupled set of equations has to be solved self-consistently for the density of
every baryon. In model 1 the coupling strength is known \emph{a priori} from the
given total baryon density but the rearrangement term in Eq.~(\ref{eq:chem1})
can only be calculated from Eq.~(\ref{eq:rearr1}) if the composition of the
neutron star matter is known. For model 2 the calculation gets even more complex
since in addition the strength of the vertices changes with the particle ratios.
Furthermore, the rearrangement self-energies of Eq.~(\ref{eq:rearr2}) differ for
each baryon multiplet and are strongly affected by the density dependence of the
vertices and the density of each particle species. Therefore, one has to include
all particles in the calculation and to check for their appearance such that the
condition of chemical equilibrium is retained. This leads to a highly nonlinear
set of equations where Eqs.~(\ref{eq:chem1})-(\ref{eq:constraint}) have to be
solved self-consistently and the calculation is far more involved than in
standard RMF theory.

\subsection{$\Lambda$ matter}
\label{ssec:NucMatLambda}

We first study the properties of strange matter for our different choices of the
density dependence of the hyperon-meson vertices. This is done by calculating
the equation of state for symmetric nuclear matter ($\rho_p = \rho_n = \frac{1}
{2}\rho_N$) with a fixed admixture $f_S$ of $\Lambda$-hyperons. The strangeness
fraction for $\Lambda N$ matter is defined as
\beq
f_S = \frac{\rho_S}{\rho_T} = \frac{\rho_{\Lambda}}{\rho_N + \rho_{\Lambda}}.
\eeq
It was pointed out in \cite{Schaffner:00} that $\Lambda N$ matter will not be
the lowest energetic state of strange matter at higher densities and for higher
strangeness fractions. A full calculation ensuring chemical equilibrium has to
include also the $\Sigma$ and the $\Xi$ since  the $\Lambda$ can be converted
via non-mesonic decay-channels into these hyperons. But for a systematic study
of the effects of the different choices of the scaling factors $R_{\sigma
\Lambda}$ and $R_{\omega \Lambda}$ for models 1 and 2 we have to restrict
ourselves to the admixture of $\Lambda$-hyperons since, as discussed in
Sec.~\ref{ssec:VertexScale}, these are the only hyperons with microscopic
derivations for the scaling factors.

\begin{figure}
\centering\epsfig{file=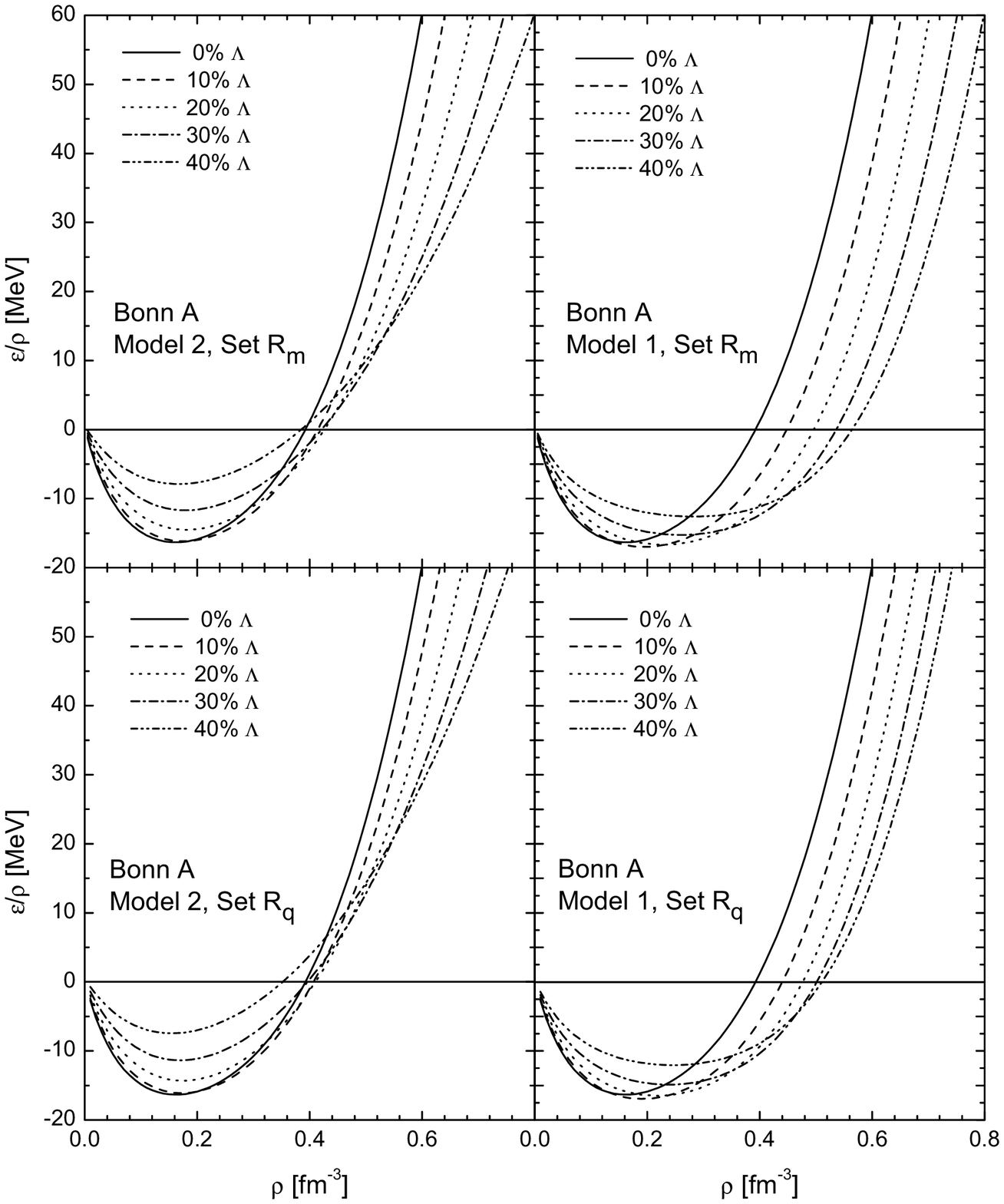,width=8.6cm}
\caption{Equation of state of symmetric nuclear matter for $\Lambda$ admixtures
from 0\% to 40\%.
Shown are results for the Bonn A NN potential for different choices of the
density dependence (model 1 and 2) and the scaling factors (set $R_q$ and $R_m$)
.}
\label{fig:ea-hypermatter}
\end{figure}

In Fig.~\ref{fig:ea-hypermatter} we show results for the equation of state
calculated with the Bonn A parameterization for different strangeness fractions.
Comparing within model 1 and 2 the differences between the set of scaling
factors $R_m$ from microscopic considerations and  the set of quark model
scaling factors $R_q$, one realizes that the equation of state is slightly
stiffer in the latter case and that the difference increases with higher
densities and higher strangeness fractions. The reason for this is that the
reduced strength of the $\Lambda$ vertices in the microscopic case decreases the
potential energy of the hyperons. The softer EoS is also in agreement with the
fact that the additional hyperon-hyperon interaction, caused by the $\sigma_s$
and $\phi$ mesons and assumed to be strongly attractive at the considered
densities \cite{Schaffner:94}, is at least in part implicitly included. At first
sight this result might be surprising since in both cases $U_{\Lambda}^{(N)}$
was adjusted to the same value. But this agreement only exists at saturation
density and for very small strangeness fractions. Since the ratio of scalar and
vector couplings changes with the density due to their medium-dependence the
equations of state differ when leaving the equilibrium density. Strictly,
identical EoS were only recovered if scalar and vector potentials scaled
proportionally to each other. This is also not the case in a RMF model with
constant couplings since the scalar and vector potentials scale differently
(partly due to nonlinear interaction terms) but the effect is less pronounced.

\begin{figure}
\centering\epsfig{file=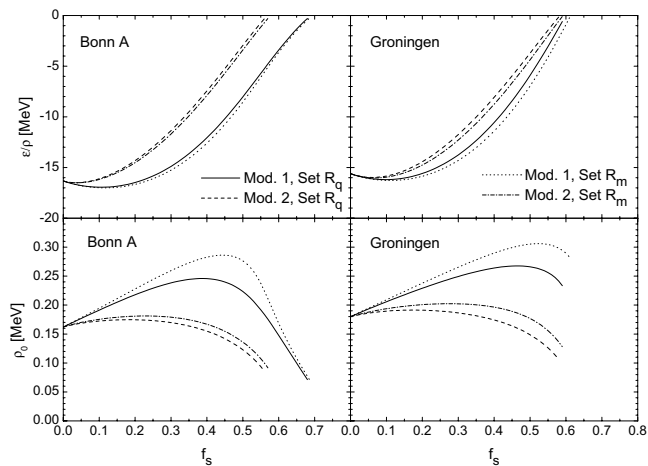,width=8.6cm}
\caption{Binding energy (top) and saturation density (bottom) as a function of
the
strangeness fraction $f_S$ for the different choices of the $\Lambda$ vertex.
Results are shown
for the Bonn A (left) and the Groningen (right) parameter set.}
\label{fig:compare-sat}
\end{figure}

\begin{figure}
\centering\epsfig{file=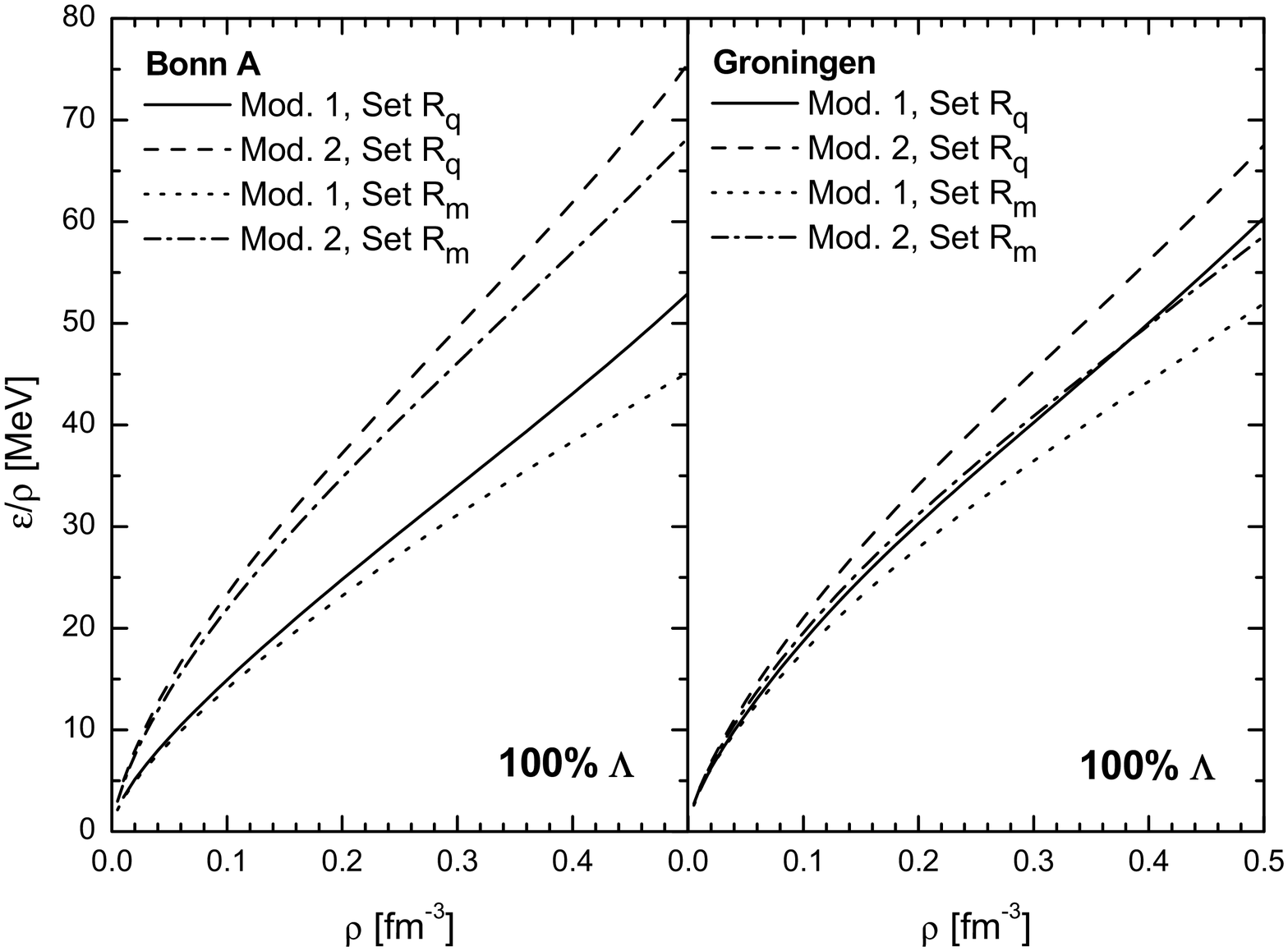,width=8.6cm}
\caption{Binding energy of pure $\Lambda$ matter as a function of the density
for the different choices of the $\Lambda$ vertex. Results are shown for the
Bonn A (left) and the Groningen (right) parameter set.}
\label{fig:100hyper}
\end{figure}

For model 1 the minimum of the EoS is shifted to higher densities and bound
stronger than in model 2 where the shift of the minima is practically
negligible. Overall, the EoS is much softer in model 1 at lower densities and
both models approach each other again at higher densities. Here, the behavior of
model 2 is in closer agreement with non-relativistic Brueckner calculations
where the position of the minimum seems to be shifted to lower densities
\cite{Stoks:99} or is relatively independent of the strangeness fraction
\cite{Schulze:98,Vidana:00b}. However, a comparison with non-relativistic
results is difficult since these do not reproduce the correct saturation density
of symmetric nuclear matter \cite{Coester:70}. This behavior is closer examined
in Fig.~\ref{fig:compare-sat} where we display the saturation density and the
binding energy at saturation density as a function of the strangeness fraction.
The Groningen parameter set exhibits a less pronounced difference between the
different models mainly due to its weaker density dependence at low densities
\cite{Hofmann:00a}. The differences between model 1 and 2 are always larger than
between the different choices $R_q$ and $R_m$ of the scaling factors. At high
strangeness fractions model 1 and 2 seem to approach each other again. This is
obvious since their density dependence is identical in the limit of pure
systems. However, the binding energy of pure $\Lambda$ matter is not identical
for model 1 and 2 as can be seen from Fig.~\ref{fig:100hyper}. The reason is
that the scaling factors have been adjusted at $f_S=0$ and differ slightly due
to the behavior of the vertices at low densities and the different choices of
the density dependence. This is seen by comparing the scaling factors of model 1
and 2 in Tab.~\ref{tab:RsRw}. Again, the difference is most pronounced for the
Bonn A potential. Results for the DD parameter set are not shown but closely
resemble the Groningen results.

This distinct difference is mainly caused by the strong density dependence of
the vertices of the Bonn A parameterization at low and very high densities. The
reason for this behavior is not the Bonn A potential itself but the polynomial
function in $k_F$ that was used for the parametrization of the self-energies.
Even though this function is very accurate around saturation density and well
suited for the calculation of finite nuclei \cite{Haddad:93} it leads to some
uncertainties when extrapolated to very low densities. Since in model 2 the
hyperon vertizes are adjusted at low densities, where due to numerical
difficulties Brueckner calculations are problematic, the extrapolation to higher
densities might involve uncertainties. The extrapolation of model 1 is safe
since the hyperon vertices are evaluated around saturation density where they
are well defined. The Groningen and the DD parameter sets avoid this problem by
choosing a rational function in $\rho$ to fit the self-energies \cite{Typel:99}
leading to a stable extrapolation to low and very high densities.

\subsection{$\beta$-stable matter}
\label{ssec:NucMatBeta}

\begin{figure}
\centering\epsfig{file=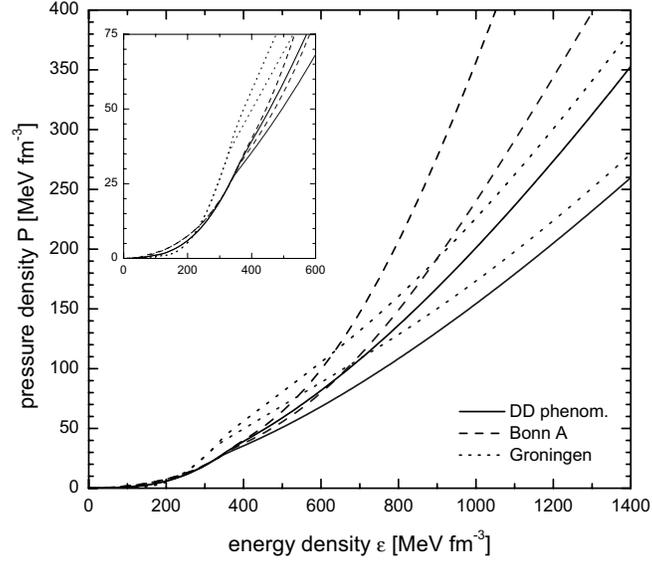,width=8.6cm}
\caption{Equation of state for neutron star matter in the DDRH model
for different density dependent interactions. Results are shown for model 1
including only the $\Lambda$ and the different scaling factors. The upper line
of each interaction corresponds to calculations with set $R_q$ and the lower one
to calculations with set $R_m$. For details see text.}
\label{fig:eps-lambda}
\end{figure}

In the following we study the composition of the equation of state for neutron
star matter as discussed in Sec.~\ref{ssec:StrangeMat}. We first examine the
case where the vertices depend on the total density (model 1) and consider only
$\Lambda$-hyperons. This allows us to compare the results with 
Sec.~\ref{ssec:NucMatLambda}. 
In Fig.~\ref{fig:eps-lambda} the pressure versus the energy
density is shown for the three different interactions. The DD parameter set has
the softest EoS, i.e. the lowest pressure at a given $\epsilon$, closely
followed by the Groningen parameter set while Bonn A is much stiffer at high
energy densities. The curves calculated with the scaling factor set $R_q$ (upper
lines) are much stiffer than the curves using the semi-microscopic set $R_m$
(lower lines). This difference is larger than expected from 
Figs.~\ref{fig:ea-hypermatter} 
and \ref{fig:compare-sat}. This is mainly due to the reduced
repulsive $\Lambda$-vector potential for set $R_m$. The appearance of the
$\Lambda$ sets in at slightly lower densities which introduces an additional
softening of the EoS by a reduction of the kinetic pressure. Remarkably, the
differences between the two choices of the scaling factors can be larger than
the differences between two parameter sets indicating that a closer examination
of the scaling factors $R_{\alpha\Lambda}$ is necessary. The properties at high
energy densities are mainly determined by the density dependence of the
vertices. Here, the Groningen and DD parameter set follow closely Brueckner
calculations. The high density extrapolation of the Bonn A parameter set leads
to an increase of the coupling strength and the contribution from the repulsive
$\omega$ field exceeds the attraction from the $\sigma$ field causing the
stiffer EoS.

\begin{figure}
\centering\epsfig{file=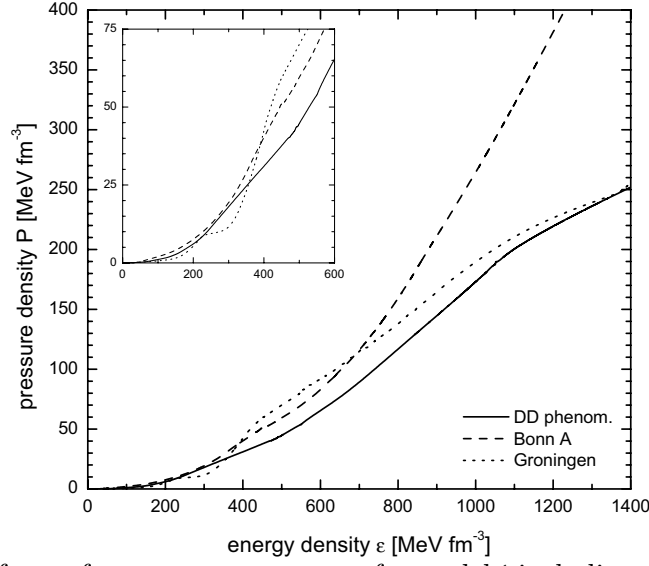,width=8.6cm}
\caption{Equation of state for neutron star matter for model 1 including all
hyperons. Results are shown for the phenomenological density dependence of
Ref.~\protect\cite{Typel:99} (solid line),
Groningen parameter set (dotted line) and Bonn A parameter set (dashed line).
The SU(6) scaling factors from set $R_q$ are used.}
\label{fig:eps-cascades}
\end{figure}

For comparison with constant coupling RMF calculations we also performed
calculations including the complete baryon octet. The scaling factors $R_{\alpha
Y}$ were determined in the standard approach from SU(6) symmetry (set $R_q$, see
Table \ref{tab:RsRw}). Results are shown in Fig.~\ref{fig:eps-cascades}. Again,
at high densities the Groningen and DD parameter set are much softer than the
Bonn A parameter set. But their properties at low and intermediate densities are
considerably different. The Groningen EoS exhibits an extreme softening at low
densities. This is explained by the early appearance of the $\Sigma^-$. 
The strong coupling to the $\delta$ field, being only present in the
Groningen parameterization, reduces the $\Sigma^-$ effective mass $m_{\Sigma^-}
^* = M_{\Sigma^-} - \Gamma_{\sigma\Sigma^-} \Phi_{\sigma} - 2
\Gamma_{\delta\Sigma^-} \Phi_{\delta} $. In contrast the isoscalar $\Lambda$
hyperon does not couple to the isovector $\delta$ meson. Moreover, because of
isospin, the effective masses of the $\Sigma^+$ and $\Xi^0$ are increased and
their treshholds are shifted to higher density regions. This behavior is
different from \cite{Schaffner:96} where the inclusion of a constant coupling
$\delta$ meson did not noticeably change the EoS.

\begin{figure}
\centering\epsfig{file=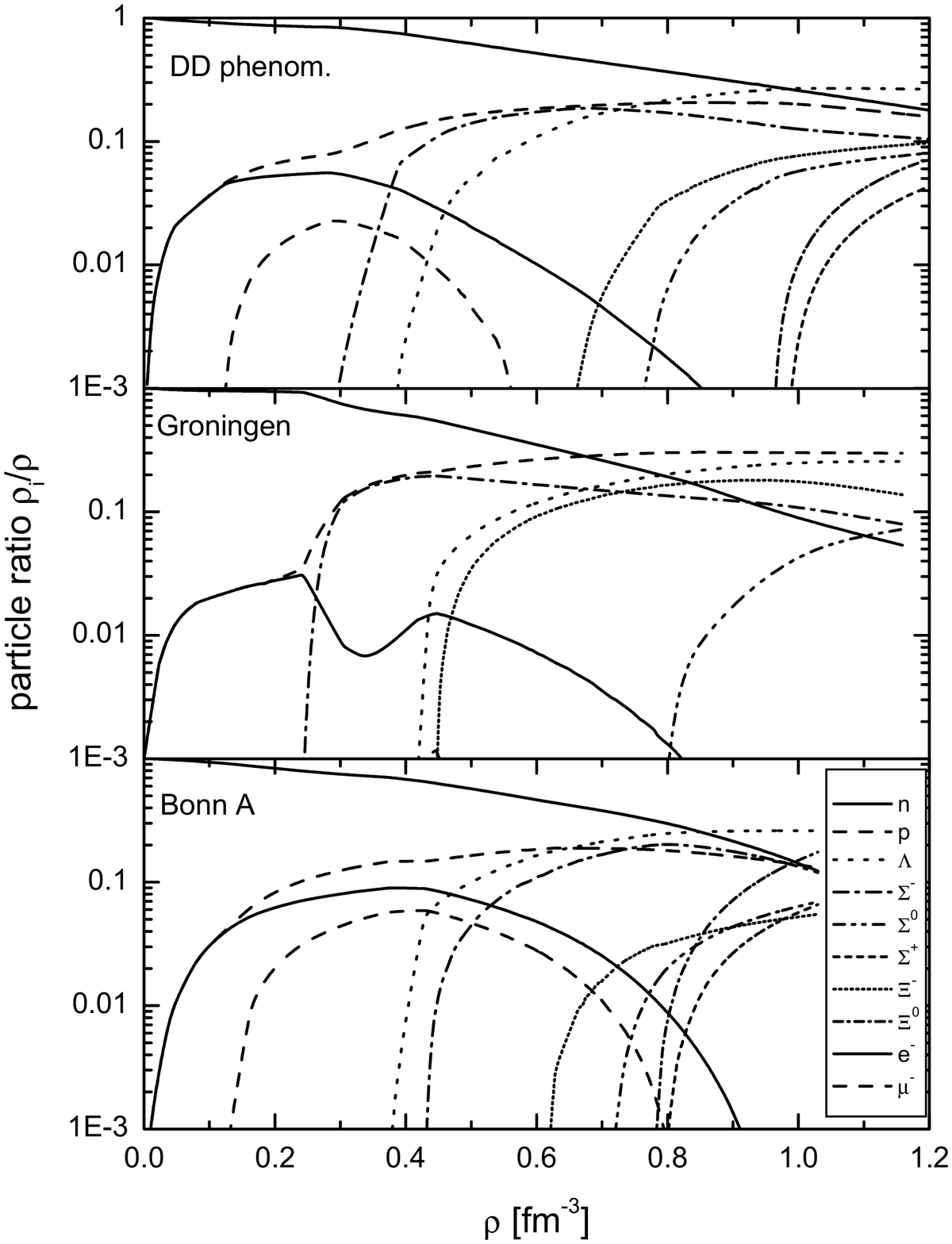,width=8.6cm}
\caption{Composition of $\beta$-stable matter in the DDRH model for the
phenomenological density dependence of Ref.~\protect\cite{Typel:99} (top),
Groningen parameter set (middle) and Bonn A parameter set (bottom). Results are
shown for model 1 including all hyperons and scaling factor set $R_q$.}
\label{fig:ratio-cascade}
\end{figure}

Results for the composition of neutron star matter for the different parameter
sets are presented in Fig.~\ref{fig:ratio-cascade}. One recognizes that the
early and strong appearance of the $\Sigma^-$ temporarily reduces the electron
fraction and strongly suppresses the myons. Clearly, this behavior opposes other
results of RMF and Brueckner calculations, e.g. \cite{Schaffner:96,Vidana:00}.
The extraction of the isovector couplings from the Groningen potential leads to
relatively large values indicating a slight increase of the $\delta$ coupling
above saturation density. This might cause the observed behavior and could be
fixed by adjusting the scaling factor $R_{\delta\Sigma^-}$ to the strength of
the $\Sigma$ isovector potential. Since experimental data is not available we
used $R_{\Sigma\delta}=R_{\rho\Sigma}=2$ from SU(6) symmetry instead. Further
investigations are necessary.

The composition of neutron star matter for the DD parameter set resembles the
outcome of non-relativistic Brueckner calculations \cite{Baldo:00,Vidana:00}
with the $\Sigma^-$ appearing before the $\Lambda$. For the Bonn A parameter set
the $\Lambda$ and $\Sigma^-$ appear at relatively high densities of about $2.5-
3\rho_0$ which also explains the stiffer equation of state. At higher densities
when the other hyperons are present the number of hyperons exceeds the number of
neutrons. This behavior is observed for all parameter sets with the strongest
effect for the Groningen parameter set. The same result was also found in other
models, e.g. \cite{Glendenning:82,Schaffner:96}.

As mentioned in Sec.~\ref{sec:Intro} one problem of RMF calculations with
coupling constants fitted to properties of finite nuclei is the appearance of
negative effective masses in neutron star matter at densities of $\rho\approx
0.5 - 1$ fm$^{-3}$ \cite{Schaffner:96}. This is mainly caused by the small
effective mass at saturation density that is needed in order to reproduce the
observed spin-orbit splitting in finite nuclei \cite{Reinhard:88}. This causes a
rather stiff equation of state and confirms that the extrapolation of
phenomenological parameter sets to high densities can be problematic. In the
DDRH approach negative effective masses appear only at much higher densities
where the validity of the model is already questionable because baryons and
mesons will cease to be the relevant degrees of freedom. We therefore stop the
calculation when negative effective masses appear and do not attempt to
extrapolate to higher densities as in \cite{Schaffner:96,Huber:98}. For the Bonn
A parameter set $m^*$ gets negative around $\rho\approx 1-1.2$ fm$^{-3}\approx6-
8\rho_0$ being mainly caused by the mentioned increase of the $\sigma$ coupling
strength at high densities. For the Groningen parameter set the neutron
effective mass gets negative at around $\rho\approx 1.2$ fm$^{-3}$. The reason
is here the additional reduction of $m^*_n$ by the coupling to the $\delta$
meson. For the DD parameter set $m^*$ is still positive for the highest
considered densities of $\rho = 1.5$ fm$^{-3}$. The different behavior of the
DDRH model compared to constant coupling models originates in the density
dependence of the vertices and is mainly caused by the fact that the vertex
density dependence takes into account information from wider density range as
found in finite nuclei and allows a safer extrapolation to higher densities. The
medium dependence of the vertices implicitly includes higher order correction
terms that can be partly identified with the nonlinear meson self-interactions
of RMF models. Also, this explains why models with nonlinear vector self-
interactions \cite{Bodmer:91,Sugahara:94} follow more closely DB calculations
avoiding negative effective masses.

\begin{figure}
\centering\epsfig{file=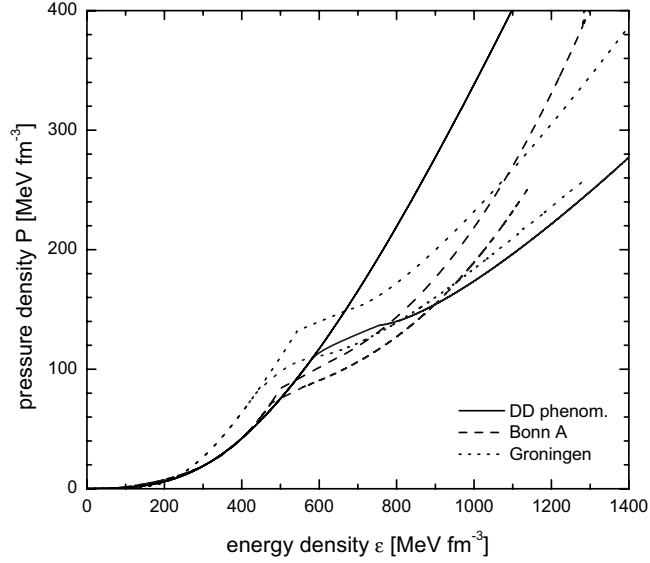,width=8.6cm}
\caption{Same as Fig.~\ref{fig:eps-lambda}, but results are shown for model 2.}
\label{fig:eps-strange}
\end{figure}

\begin{figure}
\centering\epsfig{file=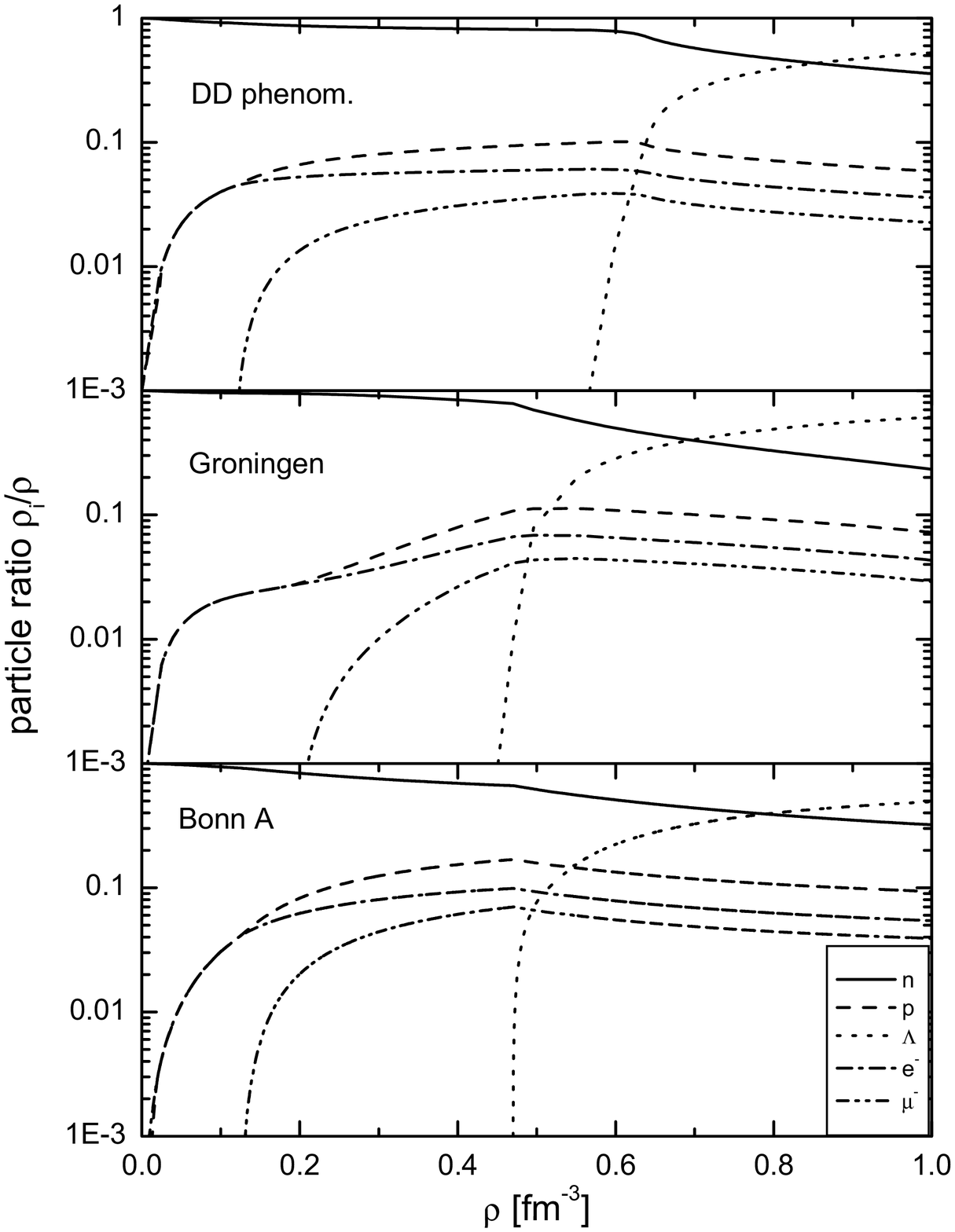,width=8.6cm}
\caption{Same as Fig.~\ref{fig:ratio-cascade},
but the density dependence is implemented with model 2. The scaling factor set
$R_m$ was used.}
\label{fig:ratio-strange}
\end{figure}

Finally, we turn to the discussion of the results obtained with model 2. In
Fig.~\ref{fig:eps-strange} the equation of state for the different density
dependent interactions is shown and Fig.~\ref{fig:ratio-strange} displays the
corresponding particle distributions. Compared to model 1 the hyperons appear at
very high densities of about $3\rho_0$. In addition their ratio increases very
fast and around $3-4\rho_0$ the number of $\Lambda$s already exceeds the number
of neutrons. This leads to a stiffer EoS at low densities and a strong softening
above $\epsilon = 500$ MeV\,fm$^{-3}$. The suppression of the hyperons in model
2 is a direct consequence of the different choice of the density dependence. The
different rearrangement self-energies for each baryon multiplet induce a
relative shift in the chemical potentials that impedes the appearance of the
hyperons. The differences between the choices of scaling factors $R_{\alpha
\Lambda}$ are the least pronounced for the Bonn A parameter set having the
strongest density dependence and the most pronounced for the DD parameter set
having the weakest density dependence. Here, the $\Lambda$ does not appear for
the densities considered if the scaling factor set $R_q$ is chosen. This leads
to the stiffest equation of state. For the semi-microscopic set $R_m$ all
interactions have very similar equations of state approaching asymptotically the
results of model 1 at high densities. Using set $R_q$ from SU(6) symmetry and
including all octet hyperons we found that the appearance of the $\Sigma$ and
the $\Xi$ is shifted to even higher densities ($\rho > 0.7$ fm$^{-3}$) or is
completely suppressed for all interactions and does not noticeably modify the
EoS.

We conclude that the composition of neutron star matter strongly depends on the
density dependence of the interaction. In practice this is problematic in model
2 since the low density behavior of the vertices has to be extrapolated from DB
calculations and introduces additional uncertainties. On the other hand, there
are indications that model 2 resembles the outcome of Brueckner calculations.
Improvements of DB calculations in the low density region might resolve these
problems. Again, we remark that we expect that a realistic density dependence is
a mixture of model 1 and 2 and has to be determined from DB calculations of
strange matter.

\section{Neutron Stars}
\label{sec:NeuStars}

The condition of hydrodynamical equilibrium inside a neutron star determines the
connection between experimentally observable properties like the mass and
possibly the radius of a neutron star and the theoretical equation of state of
neutron star matter. In this work we only consider static spherically symmetric
neutron stars. The condition of hydrodynamical equilibrium in general relativity
is expressed by the Tolman-Oppenheimer-Volkoff (TOV) differential equations
\cite{Oppenheimer:39}
\beqa
\frac{dm(r)}{dr} & = & 4\pi r^2\epsilon(r) \\
\frac{dP(r)}{dr} & = & -\frac{\left[ \epsilon(r)+ P(r) \right]\left[ m(r)+ 4\pi
P(r) r^3 \right]} {r^2\left[ 1-\frac{2m(r)}{r} \right]},
\eeqa
where $r$ is the radial distance from the origin of the neutron star and $m(r)$
is the mass contained in a sphere of radius $r$ inside the star. The
gravitational constant $G$ has been set to 1 for simplicity. Pressure $P(r)$,
energy $\epsilon(r)$ and mass $m(r)$ distributions can be calculated for a given
pressure-energy relation $P(\epsilon)$ as discussed in section \ref{sec:BetaMat}
. The differential equations are solved by integrating from the center, starting
with a central pressure $P_c$, until the surface (radius $r=R$, $P(R)=0$) of the
neutron star is reached. The radius $R$ of the star and its gravitational mass
are related by $M=m(R)=4\pi\int_0^R dr r^2 \epsilon(r)$. Varying the central
pressure one finds a mass-radius relation $M(R)$ describing a family of neutron
stars that will depend on the choice of the equation of state.

Since it is a well known fact that Brueckner theory breaks down at very low
densities one cannot rely solely on an EoS derived from the $NN$ interaction for
the description of neutron stars. Furthermore, in the surface region of neutron
stars a so called crust of sub-nuclear density (where $\rho$ is much lower than
the normal nuclear matter saturation density $\rho_0$) exists consisting of
atoms and nuclei and having its own specific degrees of freedom. Even though the
crust contributes only about $1\%$ to the total mass and does not strongly
affect the maximum mass of the star, its thickness is about $10\%$ of the radius
and influences the size of very light stars.

For the above reasons, we use the Baym-Pethick-Sutherland (BPS) EoS 
\cite{BPS:73} 
for sub-nuclear densities $\rho < 0.001$ fm$^{-3}$ and results of Negele and
Vautherin for densities of $\rho > 0.001$ fm$^{-3}$ \cite{Negele:73}. The
transition density to the DDRH EoS is defined by the intersection of both EoS.
One finds values between $0.3\,\rho_0$ and $0.7\,\rho_0$ depending on the
different interactions.

\begin{figure}
\centering\epsfig{file=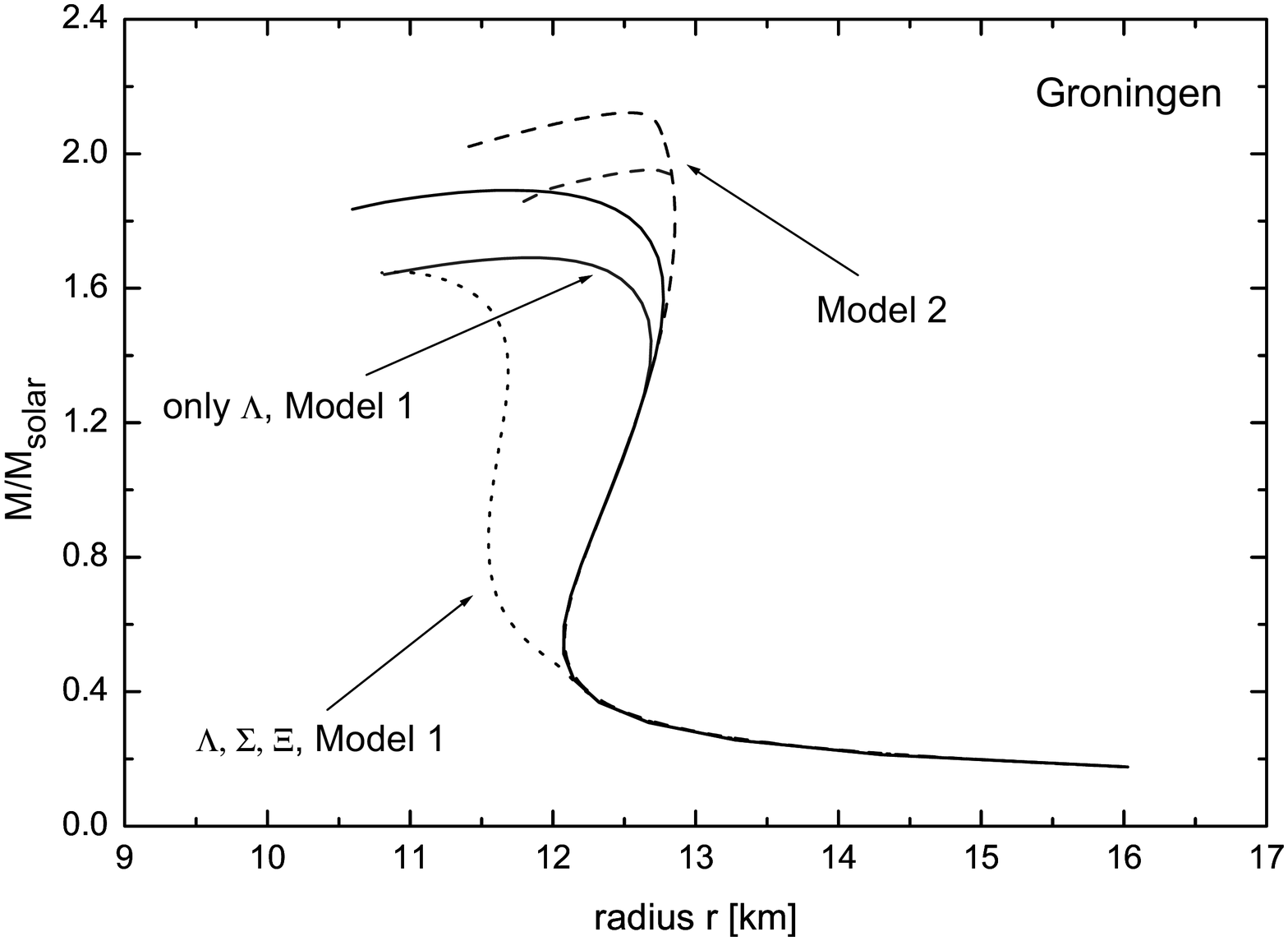,width=8.6cm}
\caption{Neutron star mass as a function of radius in the DDRH model for the
Groningen parameter set. Results for model 1 including only $\Lambda$ (solid
lines), model 1 including all hyperons (dotted lines) and model 2 (dashed lines)
are compared. The upper line of each model corresponds to calculations with set
$R_q$ the lower one to calculations with set $R_m$. For details see text.}
\label{fig:mr-groningen}
\end{figure}

In Fig.~\ref{fig:mr-groningen} mass-radius relations calculated with the
equation of state of the Groningen parameter set are shown. For model 1 we find
maximum masses of $M_{\text{max}}= 1.69 M_\odot$ for the semi-microscopic
scaling factor set $R_m$ and $M_{\text{max}}= 1.89 M_\odot$ for the SU(6) set
$R_q$. Including the full set of octet hyperons into the EoS the mass is further
reduced to $M_{\text{max}}= 1.65 M_\odot$. The maximum density in the center of
the star is similar in all three models and reaches $\rho_c \simeq 0.91 - 0.99$
fm$^{-3}$ which corresponds to $5-6\rho_0$. Including all hyperons the minimum
radius of the neutron star decreases from $R_{\text{min}}\simeq 11.7$ km to
$R_{\text{min}}\simeq 10.8$ km and the radius of neutron stars with the typical
experimentally observed mass of $M = 1.4 M_\odot$ is reduced from $R=12.7$ km to
$R=11.7$ km. The smaller radii are caused by the early occurrence of the
$\Sigma^-$ and the softening of the EoS at low densities that was discussed in
Sec.~\ref{ssec:NucMatBeta}. Also, the star mainly consists of hyperons as can be
seen from the particle composition of Fig.~\ref{fig:ratio-cascade}. Model 2
leads to significantly higher maximum masses of $M_{\text{max}}= 1.95 M_\odot$
($R_m$) respectively $M_{\text{max}}= 2.12 M_\odot$ ($R_q$) because of the
appearance of the hyperons only at higher densities. Since, in addition, the
maximum central density is only about $\rho_c \simeq 0.73$ fm$^{-3}$ the star is
mainly composed of nucleons.

From these results the density dependence of model 1 seems to be favored because
of the closer agreement with other calculations of neutron star masses. The big
difference between the maximum masses for the two sets $R_q$ resp. $R_m$
emphasizes the sensitivity of the model to the choice of the scaling factors
$R_{\alpha Y}$. The effect is as large as the inclusion of additional hyperons
and should also be examined in RMF calculations with density independent
coupling constants.

\begin{figure}
\centering\epsfig{file=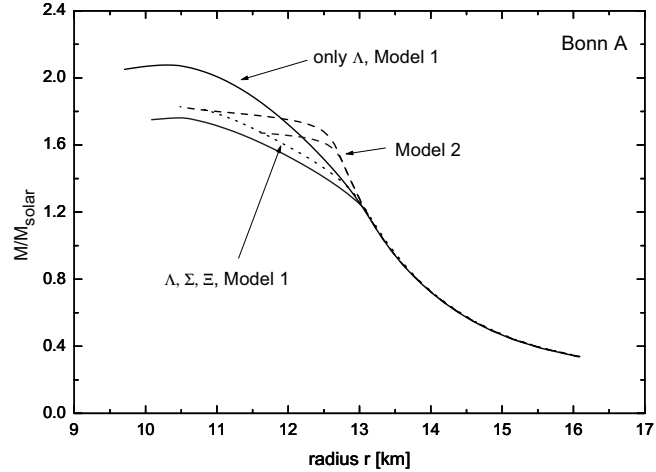,width=8.6cm}
\caption{Same as Fig.~\ref{fig:mr-groningen} but for the Bonn A parameter set.}
\label{fig:mr-bonn}
\end{figure}

For the Bonn A parameter set we find comparable results as is seen from
Fig.~\ref{fig:mr-bonn}. The maximum masses for model 1 of $M_{\text{max}}= 1.76
M_\odot$ ($R_m$) resp. $M_{\text{max}}= 2.07 M_\odot$ ($R_q$) (only $\Lambda$)
and the central density $\rho_c \simeq 1.15$ fm$^{-3}$ are higher than for the
Groningen parameter set due to the stiffer equation of state. Including the full
set of octet hyperons the maximum mass is not reached before the effective mass
gets negative at about $\rho \simeq 1.02$ fm$^{-3}$ where a mass of $M=1.81
M_\odot$ is found. The same is observed for model 2 where the effective mass
gets negative at about the same density indicating maximum masses higher than
$1.7 M_\odot$ ($R_m$) resp. $1.8 M_\odot$ ($R_q$). Despite these differences the
radii of neutron stars with $1.4 M_\odot$ are all in the range of $R=12.5-12.8$
km and in close agreement with the results for the Groningen parameter set.

\begin{figure}
\centering\epsfig{file=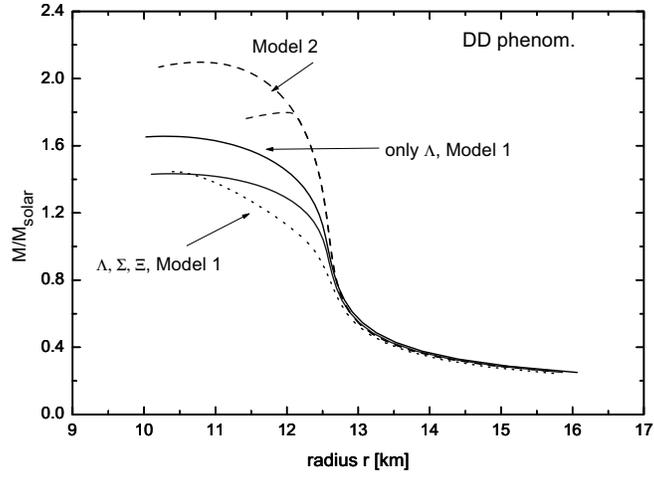,width=8.6cm}
\caption{Same as Fig.~\ref{fig:mr-groningen} but for the phenomenological
density dependence of Ref.~\protect\cite{Typel:99}.}
\label{fig:mr-typel}
\end{figure}

Results for the DD parameter set are displayed in Fig.~\ref{fig:mr-typel}.
Having the softest equation of state at high densities, this interaction also
exhibits the smallest maximum masses of $M_{\text{max}}= 1.43 M_\odot$ ($R_m$)
resp. $M_{\text{max}}= 1.66 M_\odot$ ($R_q$) (Model 1, only $\Lambda$) and
$M_{\text{max}}= 1.44 M_\odot$ (Model 1, all hyperons). For model 2 the masses
are again higher ($M_{\text{max}}= 1.77 M_\odot$ resp. $M_{\text{max}}= 2.08
M_\odot$) since the stars are mainly composed of nucleons. The DD parameter set
reaches central densities up to $\rho_c \simeq 1.3$ fm$^{-3}$ in model 1 and
$\rho_c \simeq 1.0$ fm$^{-3}$ in model 2. The soft equation of state reduces
also the radius of the neutron stars. One finds a minimum radius of
$R_{\text{min}}\simeq 10.3$ km and a radius of $R \simeq 10.8$ km for a mass of
$1.4 M_\odot$ if all hyperons are included.

Even though the three different parameter sets and the different choices of the
density dependence give different predictions for the maximum mass of a neutron
star the radius of a typical neutron star with a mass of $M=1.4 M_\odot$ is in
all models expected to be about $12.5$ km while the minimum radius can get as
small as $10.3$ km. This is explained by the fact that the radii are mainly
influenced by the low to intermediate density behavior of the EoS which is
similar in all considered models. On the other hand, the maximum masses are
strongly influenced by the high density behavior being quite different for the
investigated interactions and models 1 and 2. Our results are compatible with
non-relativistic Brueckner calculations \cite{Baldo:00,Vidana:00,Akmal:98} and
relativistic Hartree-Fock calculations (fitted to the Bonn A NN potential)
\cite{Huber:98} that predict radii of $10-12$ km and also possess a rather soft
equation of state. In contrast, most relativistic mean-field models with density
independent coupling constants find radii of $13-15$ km, e.g. 
\cite{Glendenning:82,Schertler:00}. 
Our maximum neutron star masses of $1.44-1.8 M_\odot$ agree
with the outcome of different RMF calculations while non-relativistic Brueckner
calculations favor values of $\sim 1.3 M_\odot$. This indicates that the non-
relativistic EoS might be too soft at higher densities where relativistic
effects become more important. The high value of $\sim 1.8 M_\odot$ found for
the Bonn A potential is probably caused by the inaccuracies from the polynomial
extrapolation of the density dependent vertices to the high density region.
Therefore, we favor the results of the Groningen and DD parameter set. We also
performed calculations with a purely hadronic EoS to examine the total effect of
the hyperons on the EoS. For the DD interaction a maximum mass of $2.05 M_\odot$
was found while the Groningen and Bonn A parameter sets have masses of $2.35
M_\odot$ resp.~$2.45 M_\odot$. The observed reduction $\Delta M \sim 0.5-
0.7M_\odot$ of the maximum mass caused by the inclusion of the hyperons is
comparable to other models \cite{Balberg:97,Baldo:00,Vidana:00}.

\section{Summary and conclusion}
\label{sec:Summary}

We have extended the DDRH model to the strangeness sector including the full set
of SU(3)$_f$ octet baryons. Using realistic density dependent interactions
derived from DB calculations we examined the properties of neutron star matter.
Starting with a medium-dependent NN interaction that is parameterized by density
dependent meson-nucleon vertices we found that the extension of DB results from
nuclear matter to hypermatter is not unique. The structure of the DB interaction
strongly indicates that the ratio $R_{\alpha Y}$ of the nucleon and hyperon in-
medium vertices and self-energies is mainly determined by the ratio of the
corresponding free-space coupling constants and is only weakly affected by the
background medium. We developed two different models to describe the connection
of the density dependence of the baryon vertices to the surrounding medium. In
model 1 the vertices depend on the total baryon density and  are not influenced
by the strangeness asymmetry of the medium. This ensures that the relative
strength of the vertices remains constant and follows closely the behavior of
RMF calculations with constant couplings. An extreme assumption in this model is
that the vertices are influenced equally from all baryons. However, in first
order hyperons and nucleons should be independent of each other. This was
investigated in model 2 that assumes a dependence of the vertices only on the
baryons within the corresponding SU(3)$_f$ multiplet. While this leads to a more
realistic description of slightly asymmetric systems, the model becomes
problematic at very high asymmetries since it completely neglects the higher
order effects from of the other baryon multiplets. In addition, in such systems
the low density behavior turns out to be important which is not well determined
from Brueckner calculations. We examined the properties of $N\Lambda$ matter and
found differences between the two models mainly at low densities and strangeness
fractions $f_S$ that are nearly vanishing at high densities and values of $f_S$
close to one. The conclusion is that a realistic medium dependence of the in-
medium vertices should be a mixture of both models including the asymmetry and
the total density. Such a description is pending until DB calculations for the
full strangeness octet will be available.

Application of the DDRH theory to weak $\beta$-stable matter leads to more
complex condition for chemical equilibrium than in mean-field calculations with
constant coupling constants. The occurrence of rearrangement contributions to
the chemical potential connects all particle densities strongly to each other
which complicates the practical calculation considerably.

An important aspect in our calculations is the value of the scaling factors
$R_{\alpha Y}$. Microscopic calculations of the $\sigma\Lambda$ vertex show a
strong deviation from the quark model value. Choosing either $R_{\sigma\Lambda}=
0.49$ from $N\Lambda$ T-matrix results (scaling factor set $R_m$) or
$R_{\omega\Lambda}=2/3$ from SU(6) symmetry considerations (set $R_q$) and
fixing the remaining scaling factors to the same experimental potential depth of
the $\Lambda$ we found strong differences in the equation of state. The smaller
microscopic values lead to a much softer EoS and reduce the calculated maximum
neutron star masses by 10-15\%. This behavior even though enhanced by the
density dependence of the interaction will also be visible for parameter sets
with density independent coupling constants and should be closely examined.
Microscopic calculations for the $\Sigma$ and $\Xi$ vertices are not available
restricting full scale neutron star calculations to the SU(6) symmetry values.
But one has to review if these values are realistic and study effects of their
modification on neutron star properties \cite{Glendenning:91,Huber:99}.

We remark that our calculations did not examine the influence of the hidden-
strangeness mesons $\sigma_s$ and $\phi$ explicitly that are assumed to cause a
highly attractive hyperon-hyperon interaction at intermediate densities and
might lead to a small reduction of the neutron star radii. On the other hand,
the couplings of these mesons posses some uncertainties that cannot be reliably
fixed since experimental data are missing. In addition they are in part
implicitly included in the microscopic scaling factors.

An examination of neutron star properties favors model 1 as choice for highly
asymmetric dense systems. While model 2 predicts too high maximum masses and a
suppression of the hyperons, the results for model 1 are in close agreement with
other calculations. For all examined density dependent interactions we find
radii of $10-13$ km for neutron stars with masses above $1.4 M_\odot$. This is
also observed in non-relativistic Brueckner calculations while RMF calculations
with phenomenological interactions usually favor larger radii. Our maximum
masses of $M_{\text{max}}= 1.65 M_\odot$ for the Groningen parameter set and
$M_{\text{max}}= 1.44 M_\odot$ for the DD parameter set are confirmed by other
RMF calculations and are slightly higher than in Brueckner calculations.

We conclude that the DDRH model allows a consistent calculation of strange
matter and neutron stars and yields results that are comparable with other
models. It incorporates the properties of the DB model using microscopic
interactions at various densities as input. The extrapolation to higher
densities is more constrained than for phenomenological RMF calculations that
use only information from the limited density range of finite nuclei for the
determination of their parameters. Improvements of the results, a more realistic
density dependence of the hyperon-meson vertices and more restricted predictions
will be possible if results from upcoming DB calculations for the SU(3)$_f$
baryon octet are available.

\section*{Acknowledgement}
We would like to thank C.~Greiner and K.~Schertler for interesting discussions
and helpful comments. This work was supported in part by DFG (Contract
No.~Le439/4-3), GSI
Darmstadt, and BMBF.




\begin{table}
\begin{center}
\begin{tabular}{lcc|cccc}
	& \multicolumn{2}{c}{$R_m$} & \multicolumn{4}{c}{$R_q$} \\
  model 1
  & $R_{\sigma\Lambda}$ & $R_{\omega\Lambda}$
  & $R_{\sigma\Lambda,\Sigma}$ & $R_{\omega\Lambda,\Sigma}$
  & $R_{\sigma\Xi}$ & $R_{\omega\Xi}$ \\
  \hline
  Bonn A     & 0.49 & 0.4935 & 0.6225 & 2/3 & 0.3618 & 1/3 \\
  Groningen  & 0.49 & 0.5217 & 0.6061 & 2/3 & 0.3343 & 1/3 \\
  DD phenom. & 0.49 & 0.5100 & 0.6170 & 2/3 & 0.3421 & 1/3 \\
  \hline \hline
  	& \multicolumn{2}{c}{$R_m$} & \multicolumn{4}{c}{$R_q$} \\
  model 2
  & $R_{\sigma\Lambda}$ & $R_{\omega\Lambda}$
  & $R_{\sigma\Lambda,\Sigma}$ & $R_{\omega\Lambda,\Sigma}$
  & $R_{\sigma\Xi}$ & $R_{\omega\Xi}$ \\
  \hline
  Bonn A     & 0.49 & 0.5690 & 0.5690 & 2/3 & 0.3040 & 1/3 \\
  Groningen  & 0.49 & 0.5387 & 0.5921 & 2/3 & 0.3222 & 1/3 \\
  DD phenom. & 0.49 & 0.5251 & 0.6061 & 2/3 & 0.3287 & 1/3 \\
\end{tabular}
\caption{Scaling factor sets $R_q$ and $R_m$ for the $\sigma$ and $\omega$
hyperon-meson vertices of model 1 and 2. For details see text.}
\label{tab:RsRw}
\end{center}
\end{table}

\end{document}